\def\spttt{spacetime~}
\def\m{\eqno}
\def\di{\partial}
\documentclass[12pt]{article}
\usepackage{graphicx}
\begin{document}

\title{Disk Sources for Conformastationary Metrics}
\author{Katz J$^{1,2,3}$, Bi\v{c}\'{a}k J$^{3,2}$, Lynden-Bell D$^{2,4,3,5}$.}
\maketitle
\centerline{Racah Institute of Physics$^1$, The Hebrew University,
Jerusalem 91904, Israel;}
\centerline{Institute of Astronomy$^2$, The Observatories, Cambridge, CB3 0HA, United Kingdom;}
\centerline{Institute of Theoretical Physics, Charles University$^3$,
V.Hole\v{s}ovi\v{c}k\'{a}ch 2, 18000 Prague 8, Czech Republic;}
\centerline{Clare College$^4$, Cambridge \& Queens University$^5$, Belfast
, United Kingdom.}

\begin{abstract}

Conformastationary metrics -- those of the form $$ds^2 = f \left
(dt-{\cal A}_k dx^k \right )^2 - f^{-1} \left( dx^2 + dy^2 + dz^2 \right)$$
have been derived by Perjes and by Israel \& Wilson as source-free
solutions of the Einstein-Maxwell equations.  By analogy with the
conformastatic metrics which have charged dust sources it was assumed
that conformastationary metrics would be the external metrics of
charged dust in steady motion.  However for axially symmetric
conformastationary metrics we show that, as well as moving dust, hoop
tensions are always necessary to balance the centrifugal forces
induced by the motion.  Exact examples of conformastationary metrics
with disk sources are worked out in full.  Generalisations to
non-axially symmetric conformastationary metrics are indicated.

\end{abstract}

\section{Introduction}

Using considerable mathematical ingenuity Perjes (1971) and later
Israel \& Wilson (1972) showed that metrics of the form $$ds^2 = \left
(VV^* \right ) ^{-1} {\left (dt - {\cal A}_k dx^k \right )}^2 - VV^* \left
(dx^2 + dy^2 + dz^2 \right ), \eqno (1.1)$$ where $V$ is a complex
solution of Laplace's equation $$\left ({\partial ^2 \over \partial
x^2} + {\partial^2 \over \partial y^2} + {\partial ^2 \over \partial
z^2} \right ) V = 0\ , \eqno (1.2)$$ satisfy the Einstein-Maxwell
equations for the electromagnetic fields, $$ {\bf E} + i {\bf H} =
\mbox{\boldmath$\nabla$} \left (1 \over V \right ), \eqno (1.3)$$
$${\bf D} + i {\bf B} = |V|({\bf E} + i {\bf H}) + i
\mbox{\boldmath$\cal A$} \times ({\bf E} + i {\bf H})\ ,$$ whenever
$\mbox{\boldmath$\cal A$}$ is a solution of $$\mbox
{\boldmath$\nabla$} \times \mbox{\boldmath${\cal A}$} = i \left ( V \mbox
{\boldmath$\nabla$} V^* - V^* \mbox {\boldmath$\nabla$} V \right )\
. \eqno (1.4)$$ In the above $\mbox {\boldmath$\nabla$}$ stands for $$
\left ( { \partial \over \partial x}\ ,\ {\partial \over \partial y}\
, \ {\partial \over \partial z} \right ) $$ and to obtain metrics that
become flat at infinity we need $V \rightarrow 1$ there.\\

Hereafter we write $$f = \left (VV^* \right )^{-1}\ . \eqno (1.5)$$
However, except for extreme Kerr-Newman metrics (see below), such
solutions are not known to be the external solutions of any real
matter distributions.  When $\mbox{\boldmath${\cal A}$} \equiv 0 $ the
metrics are conformastatic and then we know that they are external
metrics of static charged dust with the electric and gravitational
forces in exact balance [see Synge 1960] and for the continuous case,
e.g., Lynden-Bell et al. (1999).  Most workers guessed that the
conformastationary metrics would be the metrics of moving charged dust
with those forces balancing because cylindrical conformastatic systems
would make conformastationary balanced systems when set in uniform
motion along the cylinder.  However we show here that axially
symmetrical conformastationary solutions with disk sources always need
hoop tension to balance the centrifugal forces so that moving charged
dust sources are insufficient to maintain equilibrium when the dust is
accelerated with respect to the static frame.  In such cases the
electromagnetic and gravitational forces balance but the centrifugal
forces would be unbalanced for charged dust without the introduction
of these hoop tensions.

The theory of disk sources has a long history.  Morgan \& Morgan
(1969) introduced pressureless counter-rotating disks while
Lynden-Bell \& Pineault (1978) gave a self-similar disk in real
rotation.  Lemos (1989) discussed the peculiarly interesting case of
the counter-rotating photon disk which has accelerating Minkowski
space on each side and generalised the self-similar disks to include
surface pressures.  More general solutions have been obtained by
Chamorro et al (1987) and quite a number of exact solutions with and
without any radial pressure are now known (Bi\v{c}\'{a}k et al 1993
a+b, Bi\v{c}\'{a}k and Ledvinka 1993, Pichon \& Lynden-Bell 1996, Gonz\'ales
\& Letelier 1999).  Discs can now be constructed with the full freedom
in the pressure distribution and surface density but truly rotating
disks can only be constructed for those special external metrics which
are already known solutions of Einstein's equations.  Essentially all
others need to be computed although Neugebauer and Meinel (1995) made
their calculations of the finite uniformly rotating dust disk as
analytic as possible.  A similar procedure to that of constructing the
physical disk sources of the vacuum Kerr metrics has recently been
used to find disks with rotating matter and electric currents which
are sources of Kerr-Newman fields (Ledvinka et al. 1999).  The extreme
Kerr-Newman solutions are special cases of conformastationary fields.

The wonderful simplicity of the conformastatic metrics in which
electrostatic forces balance gravity, see e.g., (Lynden-Bell et
al. 1999), encouraged us to discover whether the conformastationary
metrics had sources of similar simplicity.  We were somewhat dismayed
to find that hoop tensions were a necessity although we had gained the
insight that they are only required to balance the centrifugal
accelerations.  With that exception the electromagnetic and
gravitational forces balance with the latter including gravomagnetic
forces.

The new aspect of electromagnetic disks is that while components of
{\bf E} within the surface must be continuous across the disk the
surface components of {\bf H} have a discontinuity equal to the
surface currents in it.  The normal component of {\bf B} has to be
continuous to preserve the magnetic flux through the surface.  These
conditions are readily satisfied if one chooses a complex solution of
Laplace's equation $V \left (x,y,z \right)$ defined only in the region
$z>0$ and then continues it into $z<0$ by the definition $ V \left (
x,y, -z \right) = V^* \left (x,y,z \right ).$ With this definition the
real part of $V$ is continuous across $z=0$ although the imaginary
part is not, and the imaginary part of ${\partial V} \over \partial z$
is continuous although the real part is not.  Furthermore, $VV^*$ is
continuous.  It follows that $V$ obeys Laplace's equation (1.2)
everywhere excepting $z=0$ where the charges and currents lie, and
from (1.3) the continuity conditions on ${\bf E}$ and ${\bf B}$ are
automatically accounted for by the above continuation of $V$ to
negative $z$ values.  The discontinuities in external curvatures give
both the surface mass density and the matter currents as considered by
Bi\v{c}\'{a}k \& Ledvinka (1993) and by Pichon \& Lynden-Bell (1996);
see also Ledvinka et al. (1999).

We detail these equations below.  The symmetry of the solutions above
and below the disk already ensures that the intrinsic curvatures in
the disk metric calculated from the metrics above and below exactly
fit.  

\section{Conformastationary metrics with axial \\ symmetry} 

We consider the metric (1.1) for axially symmetric spacetimes in
$$x^\mu = \left (x^0=t, \ \ x^1=R, \ \ x^2=z, \ \ x^3=\phi \right) \rm
{coordinates. }$$ Following (1.1), ${\cal A}_k$ has then only one
non-vanishing component in these coordinates $${\cal A}_k dx^k \equiv
{\cal A}Rd\phi \ .  \eqno (2.1)$$ The metric can be written $$ ds^2 =
f \left (dt - {\cal A} Rd \phi \right )^2 - f^{-1} \gamma _{k\ell}
dx^kdx^\ell$$ $$ = f \left (dt - {\cal A} Rd \phi \right)^2 - f^{-1}
\left (dR^2 + dz^2 + R^2 d \phi ^2 \right)\ , \eqno (2.2)$$ where $f$ and
${\cal A}$ are functions of ($R$, $z$).  The components of the metric
thus are: $$g_{00} = f\ ,\ g_{11} = g_{22} = -f^{-1}\ ,\ g_{03} =
-{\cal A}Rf\ , \ g_{33} = -R^2 f^{-1} \left (1- f^2{\cal
A}^2\right )^2\ , \eqno (2.3)$$ with their inverse being $$g^{00} =
f^{-1} \left (1- f^2{\cal A}^2 \right )\ , \ g^{11} = g^{22} = -f\ , \
g^{03} = - {{1} \over {R}} f{\cal A}\ , \ g^{33} = - {{1} \over {R^2}}
f\ , \eqno (2.4) $$ and $$\sqrt {-g} = \sqrt {-{\rm det} g_{\mu \nu}}
= f^{-1}R \ . \eqno (2.5) $$ The local tetrad $h^{(\alpha)}_\mu $ used
below is the one appearing in (2.2), i.e., $$h^{\left(0\right )}_\mu =
f^{1/2} \left (1,0,0,-{\cal A}R \right),\ h^{(1)}_\mu = f ^{- 1/2}
\left (0,1,0,0 \right ),\ h^{(2)}_\mu = f^{-1/2} \left (0,0,1,0 \right
),$$ $$ h^{(3)}_\mu = f^{- 1/2} \left (0,0,0,R \right )\ . \eqno (2.6)
$$ The dual tetrad reads (the tetrad indices being shifted by
Minkowski metric) $$h^\mu_{(0)} = f^{- 1/2} (1,0,0,0), \ h^\mu_{(1)} =
f ^{1/2} (0,1,0,0), \ $$ $$h^\mu_{(2)} = f^{1/2} (0,0,1,0), \
h^\mu_{(3)} = f^{1/2} \left ({\cal A},0,0,R^{-1} \right ). \eqno (2.7)
$$ This is the orthonormal frame used by static observers who are at
rest with respect to infinity.  The zero-angular-momentum observers
whose worldlines are orthogonal to $t$ = const. hypersurfaces will use
"locally non-rotating frames" (e.g., Misner et al 1973) given by
$$e^{(0)}_\mu = f^{1/2} \left [(1-f^2 {{\cal A}^2})^{-1/2},\ 0,\ 0,\ 0
\right]\ ,$$ $$e^{(1)}_\mu = h^{(1)}_\mu \ , \ \ \ e^{(2)}_\mu = h^{(2)}_\mu
\ , $$ $$e^{(3)}_\mu = f^{-1/2} \left [{\cal A}f^2 \left (1-f^2{\cal
A}^2 \right)^{-1/2},\ 0,\ 0,\ R \left (1-f^2 { {\cal
A}^2}\right)^{1/2} \right ]\ , \eqno (2.8) $$ and $$e^{\mu }_{(0)} =
f^{-1/2} \left [(1-f^2 {{{\cal A}^2}})^{1/2} ,\ 0,\ 0,\ - {{\cal
A}}f^2R^{-1} (1-f^2 {{\cal A}^2} ) ^{-1/2}\right ]\ ,\ $$ $$e^\mu _{(1)} =
h^\mu _{(1)},\ e^\mu _{(2)} = h^\mu_{(2)},\ $$ $$e^\mu _{(3)} =
f^{1/2} \left [0,\ 0,\ 0,\ R^{-1} \left (1-f^2 { {\cal A}^2} \right)^{-1/2}
\right ]\ . \eqno (2.9) $$ The coordinate angular velocity (i.e.,
velocity relative to infinity) of a zero-angular-momentum observer is
given by $$\omega = - {{g_{03}} \over {g_{33}}} = - {{f^2{\cal A}} \over
{1-f^2{\cal A}^2}} \ \  . \eqno (2.10)$$

\section{Axially Symmetric Conformastatic Metrics with a disk at
$z=0$}

The metric of the $z=0$ hypersurface is $$d\sigma ^2 = f (dt-{\cal
A}Rd\phi )^2 -f^{-1} \left (dR^2 +R^2d\phi^2 \right) =
g_{ab}dx^adx^b,\ \eqno (3.1) $$ where indices \ $a,b = 0, 1, 3 {\rm \
and}\ f = f(R\, , \, 0), \ {\cal A}={\cal A}(R\, , \, 0).\ $ The
components \ $g_{ab}\ {\rm and}\ g^{ab}$ are given in (2.3) and (2.4)
while $$\sqrt {-{\rm det}g_{ab}} = f^{-1/2}R \ . \eqno (3.2) $$ The
unit normal vector to the hypersurface $z = 0$ is $$n^\mu = \epsilon
_nf^{1/2} (0, 0, 1, 0) \ \ \ z = 0 \ ,\ \eqno (3.3)$$ $\epsilon _n = 1
\ \ \left (\epsilon _n = -1 \right)$ if $n^\mu$\ points in the positive
(negative) $z$ direction.

We shall now consider two spacetimes ${\cal M}_1$ and ${\cal M}_2$
with the metric (2.2) in which (see 1.5) $f = \left (VV^*\right
)^{-1}$ and the complex $V$ is a solution of e.g., (1.2) or, in $(R,
z, \phi)$ coordinates, $$\Delta V = {{1} \over {R}}\partial_R \left (R
\partial _R V \right) + \partial ^2_z V = 0\ . \eqno (3.4) $$ In
${\cal M}_1$ sources of $V$ are on the $z<0$ side, in ${\cal M}_2 \ V$
is exchanged with $V^*$ and the "conjugate" sources are on the
positive side $z>0$ -- see Figure 1.  So ${\cal M}_2$ is the
complexified mirror symmetric image of ${\cal M}_1$ in $z=0$ .  The
spacetime we are interested in is the third spacetime ${\cal M}$ which
is composed of ${\cal M}_1$ for $z>0$ and ${\cal M}_2$ for $z<0$.
\begin{figure}
\begin{center}
\includegraphics[scale=1.0]{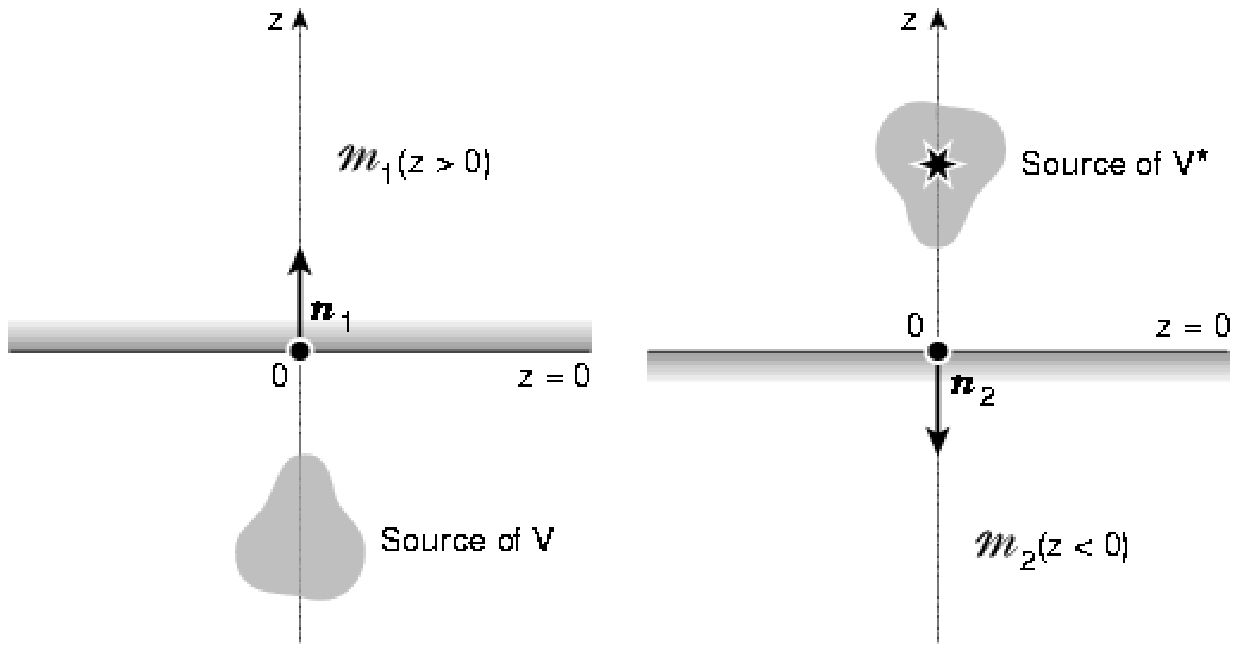}
\caption{}
\end{center}
\end{figure}

\noindent ${\cal M}$ is thus an empty space except on $z = 0$, where
there is a thin disk of matter with metric (3.1)\ .  The
energy-momentum tensor of the disk $T^\mu_{\nu} \propto \delta (z)$.
The discontinuity in the $z$ direction of $T^a_b$ defines the
energy-momentum tensor of the disk $\tau ^a_b$, more precisely, $$
\displaystyle {\lim_{\epsilon\to 0}} \int ^\epsilon
_{-\epsilon} T^a_b f^{-1/2} dz = \tau ^a_b \ , \eqno (3.5) $$ other
components are zero: $T^2_a = T^a_2 = T^2_2 = 0$ .  Thus, by
integrating Einstein's equations, $G^\nu _\mu = \kappa T^\nu _\mu \
{\rm with}\ \kappa = 8 \pi G/c^4 \ ,$ across the disk we can calculate
$\tau ^a_b$ in terms of the metric components.

The non-zero components of $\tau^{ab} = \tau ^a_cg^{c b}$ are
particularly simple if written in terms of two new quantities $$\zeta
=-\left(\partial _z f^{-1} \right)_{z=0}\ ,\ \ \chi = \left (\partial
_z{\cal A} \right )_{z=0}. \ \ \eqno (3.6)$$ We find that, $$\kappa \tau
^{00} = 2f^{1/2} \left ( \zeta + f{\cal A}\chi \right)\ ,\ \ \kappa \tau
^{03} = {{1} \over {R}} f^{3/2} \chi \ ,\ \ \kappa \tau ^{11} = \kappa
\tau ^{33} = 0\ . \eqno (3.7)$$ The projected components in the local
tetrad (2.6) are respectively $$\kappa \tau^{(0)(0)} =\kappa
\tau^{(0)}_{(0)} = 2 f^{3/2} \zeta \ \ , \ \ \kappa \tau ^{(0)(3)} =
\kappa \tau^{(3)}_{(0)} = f ^{3/2} \chi \ \ , \ \ \kappa \tau^{(2)(2)}
= \kappa \tau^{(3)(3)}=0\ . \eqno (3.8)$$

The vanishing of $\tau^{33}$ and, hence, also of the stress
$\tau^{(3)(3)}$ as measured by static observers, is the specific
feature of conformastationary metrics (2.2).  With more general
metrics of the form $$ds^2 =e^{2\nu} (dt-{\cal A}Rd \phi)^2 - e^{-2 \nu}
\left[e^{2\lambda} \left (dR^2 + dz^2 \right) + R^2d\phi ^2 \right]\ ,
\eqno (3.9)$$ in which $\lambda \left (z, R \right) \neq 0 $, as is the
case e.g., with the Kerr-Newman metrics, one obtains (see Ledvinka et
al. 1999) $$\kappa \tau _{ab} = e ^{\lambda} \left (e^{-2\lambda} g_{ab}
\right) _{z=0} \ . \eqno (3.10)$$ In our case this formula implies (3.7),
(3.8) (even though the covariant component $\tau _{33} \neq 0)$.
However in general $\tau^{33}$ is non-vanishing.  The frame components
$\tau ^{(3)(3)}$ are non-vanishing if expressed in some other than
static frame.  In particular the frames (2.8) of zero-angular momentum
observers give non-vanishing $\tau ^{(3)(3)}$ (see 4.11 below).

\section{The material properties of the disk and its motion}

If the disk is made of matter with proper rest-mass-energy surface
density \linebreak$\sigma $, surface ``hoop pressure" $\Pi $ and
angular coordinate velocity $\Omega $\ , the 3 velocity components are
$$U^a = U^0 (1,\ 0,\ \Omega),\ \ \ \Omega ={{d \phi} \over {dt}},
\eqno (4.1)$$ and $$g_{ab}U^a U^b = 1 \ .\eqno (4.2) $$ The
energy-momentum tensor is $$\tau^{1b} = 0\ , \ \ \ \tau^{ab} = (\sigma
+ \Pi)U^a U^b - \Pi g^{ab} \ \ \ a,b = 0,\ 3 \ {\rm only}. \eqno
(4.3)$$ If we compare (3.7) with (4.3), we find that $$\kappa \sigma =
f^{3/2} \left [\zeta + \sqrt {\zeta ^2 - \chi ^2} \right ]\ , \ \
\kappa \Pi = - f^{3/2} \left[ \zeta -\sqrt {\zeta ^2 - \chi ^2} \right
]\ , \eqno (4.4)$$ and $$\Omega R = {{f\chi} \over {\zeta + \sqrt
{\zeta ^2 - \chi ^2} + f{\cal A}\chi}}\ . \eqno (4.5)$$ 

Equations (4.4), (4.5) are the relationships between the matter
parameters and the geometry.  The co-moving mass energy density
$\sigma$ is positive of $$\zeta = - \left (\partial _z f^{-1}\right)
_{z=b}>0\ . \eqno (4.6)$$ It then follows that the pressure is
actually a tension $(\Pi < 0 )$ which satisfies the dominant energy
condition \ $- \Pi \leq \sigma \ $ here.  The normalisation condition
(4.2) implies $${{1} \over {(U^0)^2}} = f^{-1} \left [f^2 \left (1-R
\Omega {\cal A} \right)^2 - R^2 \Omega^2 \right] \ , \eqno (4.7)$$
which, after substituting from (4.5), gives $$U^0 = {{\zeta + \left
(\zeta ^2 - \chi ^2 \right )^{1/2} + f{\cal A} \chi} \over {\left \{
2f \left [\zeta ^2 - \chi ^2 + \zeta \left ( \zeta ^2 - \chi ^2 \right
)^{1/2} \right ] \right \}^{1/2}}} \ \ . \eqno (4.8) $$ One easily
finds the physical velocity measured by local static observers to be
$$v{\scriptscriptstyle{\rm {_{LOC}}}} = R \Omega / f \left (1 - R
\Omega {\cal A} \right )\ , \eqno (4.9)$$ whereas that measured by the
zero-angular-momentum observers reads
$$v{\scriptscriptstyle{\rm {_{ZAMO}}}} = f^{-1}R \left (\Omega - \omega
\right ) \left (1 - f^2{\cal A}^2 \right )\  , \eqno (4.10) $$ where
$\omega$\ is given by (2.10)\ .

\noindent In contrast to the vanishing stress in the $\phi$ direction
in the static frames (see 3.8) the stress does not vanish in the
zero-angular-momentum observers' frame: $$\tau
^{(3)(3)}{\scriptscriptstyle{\rm {_{ZAMO}}}} = \left (\sigma + \Pi
\right ) \left (U^0\right )^2 {\cal A}^2f^3 \left (1-f^2{\cal A}^2
\right) \left [R (\omega - \Omega)\right ]^2 + \Pi \ , \eqno (4.11)$$
where $(U^0)^2$ is given by (4.8) .\\

\noindent
The disk can only exist if $$|\chi |\leq \zeta\ , \eqno (4.12)$$
otherwise $\sigma \ , \ \Pi \ , \ {\rm and}\  \Omega $ are not real.
The limit $\chi = 0$ represents a static disk $(\Omega = 0)$ with
$\kappa \sigma = 2f^{3/2} \zeta, \ $ producing a conformastatic spacetime
on both sides.  

\section{The electromagnetic currents in the disk}     

On both sides of the disk, spacetime is free of charges and currents.
Maxwell's equations thus are $$\partial _\nu \hat F^{\mu \nu} = 0\ ,
\eqno (5.1)$$ a ``$\hat{\  }$'' means multiplication by $\sqrt {-g} =
f^{-1}R$ (see 2.5), and $F_{\mu \nu} = \partial _\mu A_\nu -
\partial_\nu A_\mu .$ According to Perjes (1971) and Israel and Wilson
(1972), and following (1.1) and (5.1), $\hat F^{k \ell}$ is of the form
$$\hat F^{k \ell} = - \eta ^{k \ell m} \partial_m \Phi \ , \eqno (5.2)$$
with $$\eta ^{k \ell m} = {1 \over \sqrt \gamma} \epsilon ^{k \ell m}
= {{1} \over {R}} \epsilon^{k \ell m} \ . \eqno (5.3)$$ $\epsilon ^{k \ell
m}$ is the alternating symbol, and $\Phi$ is some function of $R, z$ .
Both $A_0$ and $\Phi$, which, together with $g_{\mu \nu}$\ , define
$\hat F ^{\mu \nu} $ completely, are given by $$A_0 - i \Phi = f
V^{\ast} -1 \ . \eqno (5.4)$$ The spatial components of the vector
potential, $A_k$ , are here not used.  To find the charge and electric
current densities in the disk we must integrate Maxwell's equations
$$\partial_\nu \hat F^{\mu \nu} = -4 \pi \hat j_Q^\mu \eqno (5.5)$$
accross the disk where the current density $\hat j ^\mu_Q \sim \delta
(z)$.\ \ The discontinuities in the $z$ direction defines the components
$\hat i^\mu_Q$ of the surface current density
$$\displaystyle\lim_{\epsilon\to 0} \int ^\epsilon _{- \epsilon} \hat
j ^\mu _Q f ^{- 1/2} dz = \hat i ^\mu _Q \ , \eqno (5.6)$$ and the
symmetry $z \rightarrow - z$ implies that the discontinuity in the $\hat
F^{2\nu}$ components are just equal to twice $\hat F ^{2\nu}$ on
the positive side of the disk $\left (z = 0_+\right )$.  Thus, taking
account of (2.5), $$\left (F^{ \mu 2} \right )_{z=0+} = -2\pi f ^{1/2}
i ^\mu _Q \ . \eqno (5.7) $$ In terms of $\hat F ^{k \ell}$ , given by
(5.2) with (5.3), and with $$F_{0k} = - \partial _k A_0 \ , \eqno
(5.8)$$ where $A_0$ is also defined by (5.4), we find that $$\left
(\partial _z A_0 \right )_{z=0_+} = -2\pi f ^{1/2} \left (i^0_Q + \omega
R i ^3_Q \right), \eqno (5.9)$$ $${{1} \over {R}} \left (\partial _R
\Phi \right )_{z=0_+} = - 2 \pi f ^{- 1/2} i ^3 _Q \ , \eqno (5.10)$$
while $i ^2 _Q = 0$ .  Notice that if $V$ is real, $\kappa \sigma = i
^0_Q$ and the static disk is composed of charged dust in which
gravitational attraction is in equilibrium with electrostatic
repulsion.

\section{Summarising the results obtained so far} 

Given a metric of the form (2.2), with two functions $f(R,z)$ and
${\cal A}(R,z)$, a solution exists, associated with any {\it complex}
function $V (R,z)$ satisfying the Laplace equation (3.4).  \noindent
The corresponding gravitational potentials $f$ and ${\cal A}$ are defined by
(1.5) and (1.4) respectively, the electromagnetic ``potentials" $A_0$
and $\Phi$ by (5.4).  \noindent A disk at $z=0$ with $V(r, z > 0)$ and
$V^\ast (R, z < 0)$ contains matter rotating with angular velocity
$\Omega$ given by (4.5), proper mass-energy surface density $\sigma$ and hoop
tensions $- \Pi$ given by (4.4).  The surface current $i ^3_Q$ is
given by (5.10), and the charge per unit surface,  $i ^0_Q$, is
defined by (5.9).  A simple example is given later.

\section{Forces}

The equations of motion of the disk are given, [see for instance
equation (10) of Goldwirth and Katz (1995) which uses the same
notations] in terms of the normal unit vectors to $z=0$ given in
(3.3): $$\nabla _b \tau ^b_a = - \left [T^\nu _\mu {{\partial x^\mu}
\over {\partial x^a}} n_\nu \right ]^+_- = -2 T^\nu _a \ n_\nu |_{0+}
= +2f^{-1/2} T^2_a \ . \eqno (7.1)$$ $T^\nu_\mu$ is the
electromagnetic field energy-momentum tensor. The only non-trivial
equation follows for $a = 1$ for which $$\nabla_b \tau_1^b - 2
f^{-1/2} T^2_1 = - {1 \over 2} \tau ^{ab} \partial_1 g_{ab} -
2f^{-1/2} T^2_1 = 0. \eqno(7.2)$$ This equation expresses the
equilibrium of gravitational and electromagnetic forces. To see this
explicitly, consider first the $T^2_1$ term in (7.2).

From $$T^\mu_\nu = {1 \over 4\pi} \left(F^{\mu\rho} F_{\rho \nu} +{1
\over 4} \delta^\mu_\nu F^{\rho\sigma} F_{\rho
\sigma}\right)\eqno(7.3)$$ \noindent we get $$-2 f^{-1/2} T^2_1 =
-f^{-1/2} {1 \over 2\pi} \left( F^{20} F_{01} + F^{23}
F_{31}\right),\eqno(7.4)$$ or, in terms of (5.7), $$F_1 = -2f^{-1/2}
T^2_1 = i^0_Q F_{01} + i^3_Q F_{31}. \eqno(7.5)$$ This is the radial
component of the electromagnetic force, $$F_k = i^0_Q E_k + \eta_{k
\ell m} i^\ell_Q B^m, \eqno(7.6)$$ in which the components of the
electric and magnetic fields are respectively defined by $$E_k =
F_{0k} = - \partial_k A_0, \quad B^k = -{1\over 2} \eta^{k \ell m}
F_{\ell m}. \eqno(7.7)$$ $F_k$ is considered here as a 3-vector in
$\gamma_{k \ell}$ space - see (2.2). In vector notations, (7.6) has a
familiar look $${\bf F} = i^0_Q {\bf E} + {\bf i}_Q \times {\bf
B}. \eqno(7.8)$$ \noindent Consider now the $\tau^{ab} - $ term in
(7.2):  $$- {1 \over 2} \tau ^{ab} \partial _1 g_{ab} = - {1 \over 2}
\tau ^{00} \partial _1 g_{00} - \tau ^{03} \partial _1 g_{03} \
. \eqno (7.9)$$With $\tau ^{(0)}_{(0)}$ and $\tau ^{(3)}_{(0)}$ given
by (3.8) and $g_{ab}$ in (2.3), (7.9) may be rewritten as $$- {1 \over
2} \tau ^{ab} \partial _1 g_{ab} = \tau ^{(0)}_{(0)} \left (- \partial
_1 {\rm ln} f^{1/2} \right) + \tau ^{(3)}_{(0)} f \partial _1 {\cal A}_3
. \eqno (7.10)$$ This is the radial component of the gravitational force,

$${\cal F}_k = \kappa \left ( i ^0_M {\cal E} _k + \eta _{k l m} i^l_M
{\cal B}^m \right ), \eqno (7.11)$$ expressed in terms of the
gravoelectric and gravomagnetic field components defined by

$${\cal E}_k = - \partial _k {\rm ln} f^{1/2}\ , \ \ {\cal B}^m = - {1
\over 2} \eta ^{mkl}f \partial _k {\cal A}_l\ ,\ \eqno (7.12)$$ and
the matter current in the disk,

$$i ^a_M = \tau ^{(a)} _{(0)} . \eqno (7.13)$$

With (7.8) and (7.11) - (7.13), (7.2) appears clearly as the balance
of electromagnetic and gravitational forces; in vector notation, 

$$\kappa \left (i^0_M \mbox{\boldmath${\cal E}$} + {\bf i}_M \times
\mbox{\boldmath${\cal B}$} \right ) + i^0_Q {\bf E} + {\bf i}_Q \times
{\bf B} = 0 \ \ .\eqno (7.14)$$ It is worth noting that electric and
magnetic forces are not separately in equilibrium, i.e. $i ^0_M \mbox{\boldmath${\cal E}$} + i ^0 _Q {\bf E} \neq 0$ unless ${\cal A} = 0$.

Another notable point of equation (7.13) is the absence of
centrifugal and pressure forces.  These are in equilibrium by
themselves; it is best seen by considering explicitly $\tau ^{(3)(3)}$
which vanishes.  With (4.3) we find that 

$$\tau ^{(3)(3)} = f^{-1} R^{2} \tau ^{33} = f ^{-1} \left (\sigma +
\Pi \right ) \left (U^0 \right )^2 \left (\Omega R \right )^2 + \Pi =
0 \eqno (7.15)$$

In a weak gravitational field ($f\sim 1,\ \ \Pi \ll \sigma) $ and with
a slowly rotating disk ($U^0 \simeq 1$), equation (7.15) becomes the
classical condition

$$\sigma \left (\Omega R \right ) ^2 + \Pi \simeq 0 \ ,\ \eqno
(7.16)$$ for the equilibrium in a rotating narrow circular band
between the centrifugal force and the tensions that keep it from
flying apart.  Of course, under this weak-field assumption, $\tau
^{(3)(3)}
$${\scriptscriptstyle{\rm{_{ZAMO}}}}$$\approx 0$ (see 4.11) implies
(7.16) as well.

\section{A simple example}

A simple example is the following complex solution of (3.4): $$V = 1 +
{q \over r} + i { \mu (z + b) \over r ^3}\ ,\ r ^2 = R^2 + (z + b)^2
\ ,\ z > 0\ ,\ \eqno (8.1)$$ $q > 0\ ,\ b > 0\ , \ {\rm and}\ \mu \
$are constants.  This solution represents a charge and a magnetic
dipole at the same position $(0,\ -b)$.  Following (1.5), the
gravitational potential $$f = (VV^\ast)^{-1} = \left [ \left ( 1 + {q
\over r} \right )^2 + {\mu ^2 (z + b)^2 \over r^6}\right ] ^{-1}\ ,\
\eqno (8.2)$$ while, following (8.3), $$\mbox{\boldmath$\nabla$}
\times \mbox{\boldmath${\cal A}$} = i \left (V ^\ast \mbox{\boldmath$\nabla$} V -
V \mbox{\boldmath$\nabla$} V^\ast \right ) \ ,\ \eqno (8.3)$$ the
solution of which is readily found to be, in $(R,\ z,\ \phi )$
coordinates, of the form $\mbox{\boldmath${\cal A}$} = (0, 0, {\cal A}R)$ where
$${\cal A} = {\mu R \over r ^3} \left (2 + {q \over r} \right )
. \eqno (8.4)$$ Electromagnetic potentials $A_0$ and $\Phi$
follow from (5.4).  Thus, $$A_0 = 1 - f \left (1 + {q \over r}
\right )\ , \ \ \Phi = - {f \mu (z + b) \over r^3}\ ,\ \eqno (8.5)$$
and $F_{\mu \nu}$ or ${\bf E}$\ ,\ ${\bf B}$ are readily deduced from
(8.5).  The disk is associated with two functions of R defined in
(3.6); $$\zeta = - \left (\partial _z f^{-1} \right )_{z = 0} = {2b
\over r^3} \left (q + {q^2 \over r} - {\mu ^2 \over r ^3} + {3 \mu ^2
b^2 \over r ^5} \right )\ , \eqno (8.6)$$ and $$\chi = + \left
(\partial _z {\cal A} \right )_{z = 0} = - {\mu b R \over r ^5} \left
(6 + 4 {q \over r } \right )\ . \eqno (8.7)$$ Notice that $$r (z = 0)
= \sqrt {R^2 + b^2} \ . \eqno (8.8)$$ Since $b>0$ and $q>0$, the energy
condition $\zeta > 0$ is satisfied for every $b < r < \infty$\ .

Electromagnetic currents $i^0_Q \ ,\ i^3_Q$\ are, according to (5.9),
(5.10), defined in terms of $$\left (\partial _z A_0 \right )_{0+} =
f^2 \left [{qb \over r ^3} f^{-1} - \left (1 + {q \over r} \right )
\zeta \right ] = - 2 \pi f^{1/2} \left ( i^0_Q -{\cal A} R i^3_Q\right
)\ , \eqno (8.9)$$ and $$ {1 \over R} \left (\partial _R \Phi
\right)_{0+} = -f^2 {\mu b \over r ^5} \left [{3 \mu ^2 b^2 \over r
^6} - \left (1 + {q \over r} \right ) \left (3 + {q \over r} \right )
\right ]= - 2 \pi f ^{-1/2} i^3_Q \ . \eqno (8.10)$$ Spacetime becomes
flat at infinity in all directions.  In particular, for $R \rightarrow
\infty $ in the disk $(z = 0)$ $$ f \simeq 1 - {2q \over R} + {3 q^2
\over R^2}\ ,\ \ {\cal A} \simeq + {2 \mu \over R^2}\ ,\ \ \left (R
\rightarrow \infty , z = 0 \right ) \eqno (8.11)$$ $$\zeta = {2 b q
\over R^3} + {2bq^2 \over R^4} \ , \ \chi \simeq - {6 b \mu \over
R^4}\ ;\ R \rightarrow \infty \ . \eqno (8.12)$$ From this follows that \
$\kappa \sigma$ \ and $\kappa \Pi$ \ defined in (4.4) become $$\kappa
\sigma = {4bq \over R^3} \ ,\ \ \kappa \Pi = - {9 b \mu ^2 \over qR^5}
\ , \ \ R \rightarrow \infty \ ,\eqno (8.13)$$ and $$\Omega R \simeq - {3
\mu \over 2qR}\ , \ \ R \rightarrow \infty \ . \eqno (8.14)$$ Electric
charges and currents behave as follows: $$2 \pi i^0_Q = {bq \over
R^3}\ , \ \ 2 \pi i^3_Q = - {3 b \mu \over R^5} \ , \ \ R \rightarrow
\infty \ ,\eqno (8.15)$$ while
$$Gi^0_M \simeq G \sigma \simeq {1 \over 2 \pi} {bq \over R^3} \ , \ \
G i^3_M = G \sigma \Omega \simeq - {3 \over 4 \pi} {b \mu \over
R^5}\ . \eqno (8.16)$$ Thus, $$Gi^0_M \simeq i^0_Q , \eqno (8.17)$$
i.e., ``mass and charge density become equal" at great distances while
$$G i^3_Q = {1 \over 2} i^3_Q  \eqno (8.18)$$ 

Let us finally evaluate the "constants of motions" with all the G's
and c's.  According to (2.3) and following (8.2), $g_{00}=f=1-2q/r+
O(r^{-2})$. This implies, as is well known, that the total mass-energy
of this \spttt is
$$M={qc^2\over G} .
\m(8.19)
$$
 On the other hand in $x,y,z$
coordinates, the components of the gravomagnetic potential ${\cal A}_k$
 defined by (2.1)  can be calculated  with $\cal A$ given in (8.4).
Thus, in $x,y,z$ coordinates we find that 
$g_{01}={2\mu
\over r^3}y + O(r^{-3}), g_{02}=-{2\mu \over r^3}x + O(r^{-3}) $ and
$g_{03}= O(r^{-3})$. So, following for instance Carmeli (1982),[17]
equation (12), p.212, we see that the total angular momentum $J_M$ is
in the $z$ direction and is given by
$$
J_M= {\mu c^3\over G}
\m (8.20)
$$  
The electric potential  given in  (8.5)   can be written ${\cal
A}_0={q\over r}+O(r^{-2})$. Thus the total charge is
$$
Q={qc^2\over {\sqrt G}}
\m (8.21)
$$
On the other hand the components of the magnetic field which are defined
in (7.7) can be calculated from (5.3) in which $\Phi$ is given in (8.5).
At
great distance
$-B^k=B_k\simeq -\di_k \Phi$ from which follows that $B_k = {\mu \over
r^3}(\delta^{k3}- 3n^kn^3)+O(r^{-4})$ with $n^k=x^k/r$. This is formula
(44.4)
of Landau and Lifshitz (1971)[18] when  the total magnetic momentum  
$$
J_Q={\mu c^3\over {\sqrt G}}
\m (8.22)
$$
We have thus the following relations among the first integrals of this
\spttt:
$$
Q={\sqrt G}M    ~~~~,~~~~J_Q={\sqrt G}J_M 
\m (8.23)
$$
We see that the gyromagnetic ratio $J_Q/{\sqrt G}J_M=1$.

\section*{Acknowledgements}

We thank J. Ehlers for giving us the opportunity to visit the Max
Planck Institute for Gravitation in Potsdam where much of the work was
done.  Donald Lynden-Bell is supported by a PPARC Senior Fellowship.
Ji\v{r}\'{i} Bi\v{c}\'{a}k acknowledges the support from the Royal Society
and from the grant GA\v{C}R202/99/0261 of the Czech Republic.

\newpage

\end{document}